\documentclass[a4paper,11pt]{article}
\pdfoutput=1
\usepackage{jheppub}
\usepackage{amsmath,amssymb,latexsym}
\usepackage{bm}
\usepackage{hepunits}
\usepackage[svgnames]{xcolor}

\usepackage{graphicx}
\usepackage{float}
\usepackage{subfigure}

\usepackage{mathtools}
\usepackage{braket}

\usepackage{natbib}
\usepackage{graphicx}

\usepackage{shuffle} 
\usepackage{slashed} 

\usepackage{multirow}

\newcommand{\zcal}{\mathcal{Z}}
\newcommand{\scal}{\mathcal{S}}

\allowdisplaybreaks

\title{On the high-energy behavior of massive QCD amplitudes}

\author[a,b]{Guoxing Wang,}
\author[a]{Tianya Xia,}
\author[a]{Li Lin Yang,}
\author[c]{Xiaoping Ye}
\affiliation[a]{Zhejiang Institute of Modern Physics, School of Physics, Zhejiang University, Hangzhou 310027, China}
\affiliation[b]{Laboratoire de Physique Th\'eorique et Hautes Energies (LPTHE), UMR 7589, Sorbonne Universit\'e et CNRS, 4 place Jussieu, 75252 Paris Cedex 05, France}
\affiliation[c]{School of Physics, Peking University, Beijing 100871, China}
\emailAdd{wangguoxing2015@pku.edu.cn}
\emailAdd{xiatianya@zju.edu.cn}
\emailAdd{yanglilin@zju.edu.cn}
\emailAdd{yexiaoping@pku.edu.cn}

\abstract{In this note, we propose a factorization formula for gauge-theory scattering amplitudes up to two loops in the high-energy boosted limit. Our formula extends existing results in the literature by incorporating the contributions from massive loops. We derive the new ingredients in our formula using the method of regions with analytic regulators for the rapidity divergences. We verify our results with various form factors and the scattering amplitudes for top-quark pair production. Our results can be used to obtain approximate expressions for complicated two-loop massive amplitudes from simpler massless ones, and can be used to resum the mass logarithms to all orders in the coupling constant.}

\begin{document}

\maketitle

\clearpage

\section{Introduction}

It is well-known that in the high-energy limit, perturbative scattering amplitudes and form factors in gauge theories admit certain factorization properties at the leading power of $m/E$~\cite{Penin:2005eh, Mitov:2006xs, Becher:2007cu, Liu:2017axv, Blumlein:2018tmz, Engel:2018fsb, Broggio:2022htr}, where $m$ denotes the typical mass scale of internal and external particles, and $E$ denotes the typical energy scale of the hard scattering. In quantum chromodynamics (QCD), such a factorization formula up to the two-loop order has been presented in \cite{Mitov:2006xs}. However, certain contributions from bubble insertions of heavy-quark loops were not incorporated (hereafter we will refer to these contributions as the ``$n_h$-contributions''). These contributions were considered in quantum electrodynamics (QED) \cite{Becher:2007cu}, where it was shown that a soft function is needed to account for the soft photon exchanges with a heavy-fermion bubble insertion. In the context of heavy-to-light form factors, a similar factorization formula has been presented in \cite{Engel:2018fsb}. However, as far as we know, there has been no generic treatment of such contributions in non-Abelian gauge theories with an arbitrary number of massive flavors (whose masses can be different). The purpose of this note is to bridge this gap and provide a generic factorization formula that applies to any scattering amplitude up to two loops in the high-energy limit. This can be used to obtain approximate results for complicated two-loop massive amplitudes from the corresponding much simpler massless ones. With suitable renormalization group equations (RGEs), the factorization formula can also be used to resum the large logarithms $\ln^n(m/E)$ to all orders in the coupling constant.

\section{The factorization formula}

We consider the perturbative scattering amplitude of an arbitrary number of massless or massive partons, with possible colorless particles in addition. In the highly boosted limit where the energy scale of the hard-scattering process is much larger than the masses of external and internal partons, the scattering amplitude can be factorized as\footnote{If some of the parton masses are of the same order as the hard scale, the factorization formula needs to be slightly modified. See, e.g., \cite{Engel:2018fsb} in the context of heavy-to-light form factors. We don't consider this case in this work.}
\begin{equation}\label{eq:fac}
\Ket{\mathcal{M}^{\text{massive}}(\{p\},\{m\})} = \prod_i \left(\zcal_{[i]}^{(m|0)}(\{m\})\right)^{1/2}
\bm{\scal}(\{p\},\{m\}) \Ket{\mathcal{M}^{\text{massless}}(\{p\})} ,
\end{equation}
where $\{p\}$ denotes the set of all external momenta; $\{m\}$ denotes the set of all parton masses involved in this scattering process; and $i$ runs over all external partons. Note that we have suppressed the dependence of the above functions on the dimensional regulator $\epsilon$ and the renormalization scale $\mu$. The $\zcal$-factors can be expanded in $\alpha_s$ as
\begin{equation}
\zcal_{[i]}^{(m|0)} (\{m\}) = 1 + \sum_{n=1}^\infty \left( \frac{\alpha_s}{4\pi} \right)^n \, \zcal_{[i]}^{(n)} \,.
\end{equation}
The coefficients $\zcal_{[i]}^{(1)}$ and $\zcal_{[i]}^{(2)}$ have been given in \cite{Mitov:2006xs}, up to terms proportional to $n_h^1 n_l^0$, where $n_h$ is the number of heavy flavors, and $n_l$ is the number of massless flavors. The soft function $\bm{\scal}$ also contains $n_h^1 n_l^0$ contributions starting from the second order in $\alpha_s$. Its presence has been proposed in \cite{Becher:2007cu} for QED processes, but has not been extended to more general non-Abelian gauge theories. The purpose of this note is to provide these missing ingredients in the factorization formula up to order $\alpha_s^2$.

For illustration, we will take the theory as QCD with all fermions (i.e., quarks) in the fundamental representation, although our results can be easily adapted to any other gauge theories with fermions in any representations. We assume that there are several heavy flavors in the theory, labelled by the index $h$, with the mass $m_h$. Up to the second order, the soft function is given by
\begin{equation}
\label{eq:soft_operator}
\bm{\scal}(\{p\},\{m\}) = 1 + \left( \frac{\alpha_s}{4\pi} \right)^2 \sum_{\substack{i,j\\i \neq j}} \left(-\bm{T}_i \cdot \bm{T}_j\right)  \sum_{h} \scal^{(2)}(s_{ij},m_h^2) + \mathcal{O}(\alpha_s^3) \,,
\end{equation}
where $i$ and $j$ run over all external legs; $h$ runs over all heavy flavors; and
\begin{align}\label{eq:softfunc}
\scal^{(2)}(s_{ij},m_h^2) &= T_F \left( \frac{\mu^2}{m_h^2} \right)^{2\epsilon} \left( -\frac{4}{3\epsilon^2} + \frac{20}{9\epsilon} - \frac{112}{27} - \frac{4\zeta_2}{3} \right) \ln\frac{-s_{ij}}{m^2_{h}} \,.
\end{align}
The boldface $\bm{T}_i$ is the color generator for the external parton $i$ which is an operator in the color space \cite{Catani:1996jh, Catani:1996vz}. For a final-state quark or an initial-state anti-quark, $(\bm{T}^a_i)_{\alpha\beta} = t^a_{\alpha\beta}$; for a final-state anti-quark or an initial-state quark, $(\bm{T}^a_i)_{\alpha\beta} = -t^a_{\beta\alpha}$; and for a gluon, $(\bm{T}^a_i)_{bc} = -i f^{abc}$. The dot product is $\bm{T}_i \cdot \bm{T}_j \equiv \bm{T}_i^a \bm{T}_j^a$ with repeated indices summed over.

The $\zcal$-factor for massless quarks starts at $\alpha_s^2$, and is given by
\begin{equation}
\label{eq:calZq2}
\zcal_{[q]}^{(2)} = \sum_h C_F T_F \left( \frac{\mu^2}{m_h^2} \right)^{2\epsilon} \left[ \frac{2}{\epsilon^3} + \frac{8}{9\epsilon^2} - \frac{1}{\epsilon} \left( \frac{65}{27} + \frac{2\zeta_2}{3} \right)+ \frac{875}{54} + \frac{16\zeta_2}{3} - \frac{20\zeta_3}{3} \right] .
\end{equation}
The one-loop $\zcal$-factor for gluons has been given in \cite{Mitov:2006xs}, and reads
\begin{equation}
\zcal_{[g]}^{(1)} = \sum_h T_F \left( \frac{\mu^2}{m_h^2} \right)^{\epsilon} \left( -\frac{4}{3\epsilon} - \frac{2\zeta_2}{3} \epsilon + \frac{4\zeta_3}{9} \epsilon^2 \right) .
\end{equation}
The two-loop coefficient can be split into three parts:
\begin{equation}
\label{eq:zcal_g_2}
\zcal_{[g]}^{(2)} = \left( \zcal_{[g]}^{(1)} \right)^2 + \frac{4}{3\epsilon} (n_l + n_h) T_F \zcal_{[g]}^{(1)} + \sum_h \zcal_{[g]}^{(2),h} \,,
\end{equation}
where
\begin{align}\label{eq:calZg2}
\zcal_{[g]}^{(2),h} &= C_A T_F \left( \frac{\mu^2}{m_h^2} \right)^{2\epsilon} \left[ \frac{2}{\epsilon^3} + \frac{34}{9\epsilon^2} - \frac{2}{\epsilon} \left( \frac{22}{9} \ln\frac{\mu^2}{m_h^2} + \frac{64}{27} - \zeta_2 \right) \right. \nonumber
\\
&\hspace{4cm} \left. + \frac{22}{9} \ln^2 \frac{\mu^2}{m_h^2} + \frac{358}{27} + \frac{4\zeta_2}{3} - 4\zeta_3 \right] \nonumber
\\
&- C_F T_F \left( \frac{\mu^2}{m_h^2} \right)^{2\epsilon} \left( \frac{2}{\epsilon} + 15 \right) .
\end{align}
Finally, we discuss the $\zcal$-factor for a heavy flavor $Q$. The one-loop coefficient has again been given in \cite{Mitov:2006xs}. The two-loop coefficient can be split into three parts:
\begin{equation}
\zcal_{[Q]}^{(2)} = \zcal_{[Q]}^{(2),l} + \zcal_{[Q]}^{(2),Q} + \sum_{h \neq Q} \zcal_{[Q]}^{(2),h} \,,
\end{equation}
where $\zcal_{[Q]}^{(2),l}$ contains contributions from gluon loops and light-quark loops; $\zcal_{[Q]}^{(2),Q}$ denotes the contribution from a loop insertion of $Q$ itself; and $\zcal_{[Q]}^{(2),h}$ denotes the contribution from a loop insertion of heavy flavors other than $Q$. We do not give the explicit expressions for $\zcal_{[Q]}^{(1)}$ and $\zcal_{[Q]}^{(2),l}$ here since they are $n_h$-independent, and refer the readers to \cite{Mitov:2006xs}. The new ingredients $\zcal_{[Q]}^{(2),Q}$ and $\zcal_{[Q]}^{(2),h}$ are given by
\begin{align}
\zcal_{[Q]}^{(2),Q} &= C_F T_F \left[ \frac{2}{\epsilon^3} + \frac{1}{\epsilon^2} \left( \frac{4}{3} \ln\frac{\mu^2}{m_Q^2} + \frac{8}{9} \right) + \frac{1}{\epsilon} \left( \frac{4}{9} \ln\frac{\mu^2}{m_Q^2} - \frac{65}{27} - 2\zeta_2 \right) \right. \nonumber
\\
&\left. -\frac{4}{9} \ln^3\frac{\mu^2}{m_Q^2} - \frac{2}{9} \ln^2\frac{\mu^2}{m_Q^2} - \left( \frac{274}{27} + \frac{16\zeta_2}{3} \right) \ln\frac{\mu^2}{m_Q^2} + \frac{5107}{162} - \frac{70\zeta_2}{9} - \frac{4\zeta_3}{9} \right] ,
\\
\zcal_{[Q]}^{(2),h} &= \zcal_{[Q]}^{(2),Q} + C_F T_F \left\{ -\frac{8}{3\epsilon^2} H(0,x) + \frac{1}{\epsilon} \left[ \frac{64}{3} H(0,0,x) - \left( \frac{8}{9} + \frac{16}{3}\ln\frac{\mu^2}{m_Q^2} \right) H(0,x) \right] \right. \nonumber
\\
&+ \frac{332x^2}{9} - \frac{332}{9} - 128H(0,0,0,x) + \left(\frac{128}{3} \ln\frac{\mu^2}{m_Q^2} + 48x^4 - \frac{128}{9} \right) H(0,0,x) \nonumber
\\
&+  \left(-\frac{16}{3} \ln^2\frac{\mu^2}{m_Q^2} - \frac{16}{9}\ln\frac{\mu^2}{m_Q^2} + \frac{224x^2}{9} - \frac{8 \zeta_2}{3} + \frac{16}{27} \right) H(0,x) \nonumber
\\
&+ \left( 24 x^4 - \frac{220 x^3}{3} + 36 x + \frac{8}{9} \right) \zeta_2 - \frac{16 \zeta_3}{3} + \frac{32}{3} H(0,-1,0,x) - \frac{32}{3} H(0,1,0,x) \nonumber
\\
&+ \left(24 x^4-\frac{220 x^3}{9} + 12x -\frac{104}{9}\right) H(1,0,x) \nonumber
\\
&- \left(24 x^4+\frac{220 x^3}{9} - 12x -\frac{104}{9}\right) H(-1,0,x) \,,
\label{eq:ZQ2h}
\end{align}
where $x=m_h/m_Q$ and $H$ denotes the harmonic polylogarithms (HPLs) \cite{Remiddi:1999ew}. It is easy to check that $\zcal_{[Q]}^{(2),h} = \zcal_{[Q]}^{(2),Q}$ when $x=1$, i.e., $m_h = m_Q$.

\section{Derivation and validation}

In this section, we show how to derive the $n_h$-contributions to the $\zcal$-factors and the soft function, and demonstrate the validity of our factorization formula in the context of various form factors and the scattering amplitudes for $t\bar{t}$ production.

\subsection{The derivation}

In the highly-boosted limit, the scattering amplitudes can be described in the soft-collinear effective theory \cite{Bauer:2000yr, Bauer:2001yt, Beneke:2002ph}. For each external leg $i$, we introduce a light-like vector $n_i^\mu$ with space-components along the same direction of $p_i$, and a light-like vector $\bar{n}_i^\mu$ opposite to $p_i$. They satisfy
\begin{equation}
n_i^2 = \bar{n}_i^2 = 0 \,, \quad n_i \cdot \bar{n}_i = 2 \,.
\end{equation}
We define a small expansion parameter $\lambda \sim m/\sqrt{|s|}$, where $m$ is a representative mass scale, and $s$ is a representative of $s_{ij}$. The relevant momentum modes are
\begin{align}
\label{eq:modes}
\text{hard} &: k^\mu \sim \sqrt{|s|} \,, \nonumber
\\
\text{$n_i$-collinear} &: (n_i \cdot k,\, \bar{n}_i \cdot k,\, k_\perp) \sim \sqrt{|s|} \, (\lambda^2,\, 1,\, \lambda) \,, \nonumber
\\
\text{soft} &: k^\mu \sim \sqrt{|s|} \, \lambda \,. \nonumber
\end{align}
The contributions from the hard region correspond to the massless amplitudes $\mathcal{M}^{\text{massless}}$, while the collinear and soft regions produce the $\zcal$-factors and the soft function $\bm{\scal}$ in Eq.~\eqref{eq:fac}.

\begin{figure}[t!]
\centering  
\includegraphics[width=0.4\textwidth]{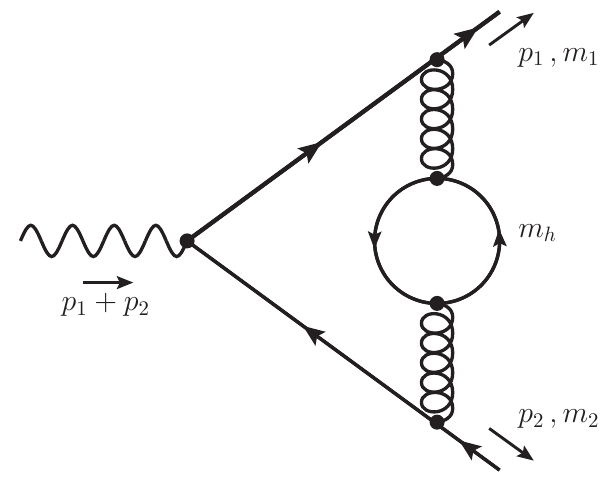}
\caption{Two-loop Feynman diagram with a massive bubble for the $q\bar{q}$-vector vertex.}
\label{fig:diagramq}
\end{figure}

We are concerned with the two-loop contributions to $\zcal$ and $\bm{\scal}$ with a massive bubble insertion, which involve the double-collinear integrals and the double-soft integrals, depicted in Fig.~\ref{fig:diagramq}. Note that without the massive bubble, the relevant integrals are well-defined, which give rise to the non-$n_h$ terms in the $\zcal$-factors (the soft integrals without the massive bubble are scaleless and vanish). However, with the massive bubble insertion, these integrals become rapidity-divergent (similar as the Sukadov problem with a massive vector boson \cite{Chiu:2007yn, Chiu:2007dg}), and one needs to introduce an additional regulator. To this end we employ the analytic regulator of \cite{Beneke:2003pa, Chiu:2007yn, Becher:2010tm, Becher:2011dz}. With this regulator, the double-soft contributions vanish, and we only need to consider the double-collinear integrals.

We will need to perform the calculations for external massless-quarks, external massive-quarks and external gluons. For convenience, we work with the two-point quark-anti-quark vector form factors and the gluon-gluon scalar form factor with outgoing external momenta $p_1$ and $p_2$. In the center-of-mass frame, the relevant light-like directions are $n^\mu = n_1^\mu = (1,0,0,1)$ and $\bar{n}^\mu = n_2^\mu = (1,0,0,-1)$. In this frame, $p_1$ and $p_2$ have no perpendicular components, and can be written as
\begin{align}
p_1^{\mu}= p_1^+\frac{\bar{n}^{\mu}}{2} + p_1^-\frac{n^{\mu}}{2} \,, \quad p_2^{\mu}= p_2^+\frac{\bar{n}^{\mu}}{2} + p_2^-\frac{n^{\mu}}{2} \,,
\end{align}
where $p_i^+ \equiv n \cdot p_i$ and $p_i^- \equiv \bar{n} \cdot p_i$. In the counting of \eqref{eq:modes}, we have $p_1^+ \sim p_2^- \sim \lambda^2\sqrt{|s|}$ and $p_1^- \sim p_2^+ \sim \sqrt{|s|}$.
By convention, we refer to the $n$-collinear modes as ``collinear'' and the $\bar{n}$-collinear modes as ``anti-collinear'', respectively. We set $p_1^2 = m_1^2$, $p_2^2 = m_2^2$ and $(p_1+p_2)^2 = s$. In the $q\bar{q}$ case, $m_1$ and $m_2$ can be zero or non-zero, while they are always zero in the $gg$ case. The mass of the internal quark is taken to be $m_h$.

We first discuss the $q\bar{q}$ case with $m_1=m_2=m_Q$. Here, the $m_Q = m_h$ case has been considered in Ref.~\cite{terHoeve:2023ehm}, and we will consider the more general case where $m_h$ and $m_Q$ can be different. The $q\bar{q}$-vector vertex can be decomposed as 
\begin{equation}
\Gamma^{\mu}(p_1,p_2) = F_{1}\big(s,m_Q^2,m_h^2\big) \, \gamma^{\mu} 
+ \frac{1}{2m_Q} F_{2}\big(s,m_Q^2,m_h^2\big) \,
i \, \sigma^{\mu\nu} (p_1+p_2)_{\nu} \,,
\end{equation}
where $\sigma^{\mu\nu} = i[\gamma^{\mu},\gamma^{\nu}]/2$. The form factors $F_i$ can be extracted by applying a set of projectors given in \cite{terHoeve:2023ehm}, and be expressed as linear combinations of scalar Feynman integrals.
 These Feynman integrals can be defined in the following generic form
\begin{multline}
\label{eq:ff_integral}
I_{\{a_i\}}
\equiv \mu^{4\epsilon}\int\frac{dk_1}{(2\pi)^d}\frac{dk_2}{(2\pi)^d}
\frac{1}{[k_1^2-m_h^2]^{a_1}}
\frac{1}{[k_2^2-m_h^2]^{a_2}}
\frac{1}{[(k_1+k_2)^2]^{a_3}}
\frac{1}{[(k_1+k_2-p_1)^2-m_1^2]^{a_4}}
\\
\times
\frac{\left(-\tilde{\mu}^2\right)^{\nu}}{[(k_1+k_2+p_2)^2-m_2^2]^{a_5+\nu}}
\frac{1}{[(k_1-p_1)^2]^{a_6}}
\frac{1}{[(k_1+p_2)^2]^{a_7}}\,,
\end{multline}
where $\nu$ is the rapidity regulator and $\tilde{\mu}$ is the corresponding scale introduced to balance the mass dimension. In our calculation, the last two propagators are actually irreducible scalar products (ISPs) with $a_6,a_7 \leq 0$. We need to calculate the above integrals in the double-collinear ($cc$) and double-anti-collinear ($\bar{c}\bar{c}$) regions (recall that the soft region integrals vanish). Note that $F_2 = 0$ at the leading power (LP) in $\lambda$, so we only need to calculate the form factor $F_1$.
The contributions from the $cc$ and $\bar{c}\bar{c}$ regions are given by
\begin{align}
F_{1,cc}^{(2),\text{bare}}\left(s,m_Q^2,m_h^2\right) &= C_FT_F\left(\frac{\mu^2}{m_h^2}\right)^{2\epsilon} \left( \frac{1}{\nu} + \ln\frac{\tilde{\mu}^2}{-s} \right) \left(\frac{4}{3\epsilon^2} - \frac{20}{9\epsilon} + \frac{112}{27} + \frac{4\zeta_2}{3} \right) \nonumber
\\
&\hspace{-3cm} +C_FT_F\left(\frac{\mu^2}{m_Q^2}\right)^{2\epsilon}\left\{ - \frac{4}{3\epsilon^3} - \frac{1}{\epsilon^2}\left[ \frac{4}{3} - \frac{8}{3} H(0,x) \right] - \frac{1}{\epsilon}\left[ \frac{8\zeta_2}{3} + \frac{56}{9} + \frac{16}{3} H(0,0,x) \right] \right. \nonumber
\\
&\hspace{-3cm}\left.  - \frac{4 \zeta(3)}{9} - \left(\frac{20 x^3}{3} - 36 x + \frac{80}{9} \right) \zeta_2 + \frac{40 x^2}{9} - \frac{2504}{81} +\left(\frac{40 x^2}{9}+\frac{8
   \zeta_2}{3}-\frac{224}{27}\right) H(0,x)  \right. \nonumber
\\
&\hspace{-3cm}\left. - \left( \frac{20 x^3}{9} - 12 x - \frac{88}{9} \right) H(-1,0,x) - \left(\frac{20x^3}{9} - 12 x + \frac{88}{9} \right) H(1,0,x) - \frac{80}{9} H(0,0,x)  \right. \nonumber
\\
&\hspace{-3cm}\left. + \frac{16}{3} H(0,-1,0,x) - \frac{16}{3} H(0,1,0,x)\right\} \,, \label{eq:F1cc}
\\
F_{1,\bar{c}\bar{c}}^{(2),\text{bare}}\left(s,m_Q^2,m_h^2\right) &= C_FT_F\left(\frac{\mu^2}{m_h^2}\right)^{2\epsilon} \left( \frac{1}{\nu} + \ln\frac{\tilde{\mu}^2}{m_h^2} \right) \left( -\frac{4}{3\epsilon^2} + \frac{20}{9\epsilon} - \frac{112}{27} - \frac{4\zeta_2}{3} \right) \nonumber
\\
&\hspace{-3cm}+C_FT_F\left(\frac{\mu^2}{m_Q^2}\right)^{2\epsilon}\left\{ \frac{2}{3\epsilon^3} - \frac{1}{\epsilon^2}\left[ \frac{28}{9} + \frac{16}{3} H(0,x) \right] - \frac{1}{\epsilon}\left[ \frac{2\zeta_2}{3}+\frac{212}{27}  -\frac{64}{9} H(0,x) \right.\right. \nonumber
\\
&\hspace{-3cm}\left.\left. - \frac{80}{3} H(0,0,x)+\right] -\frac{40\zeta_3}{9} - \left(\frac{20 x^3}{3} - 36x + \frac{32}{3}\right) \zeta_2 + \frac{40x^2}{9} - \frac{1652}{81} \right. \nonumber
\\
&\hspace{-3cm}\left. +\left(\frac{40 x^2}{9}-\frac{16\zeta_2}{3}-\frac{16}{9}\right) H(0,x) -\frac{112}{3} H(0,0,x) - \left(\frac{20 x^3}{9} - 12x - \frac{88}{9}\right) H(-1,0,x) \right. \nonumber
\\
&\hspace{-3cm}\left. - \left(\frac{20x^3}{9} - 12x + \frac{88}{9} \right) H(1,0,x) + \frac{16}{3}
   H(0,-1,0,x) - 128 H(0,0,0,x) \right. \nonumber
\\
&\hspace{-3cm}\left. - \frac{16}{3} H(0,1,0,x) \right\}\,.
\label{eq:F1cbarcbar}
\end{align}
Note that the exchange symmetry between the $cc$ and $\bar{c}\bar{c}$ regions are broken since we have only introduced the analytic regulator on the propagator associated with $m_2$. It is possible to obtain a more symmetric result by introducing two regulators $\nu_1$ and $\nu_2$ for the two propagators associated with $m_1$ and $m_2$. This makes the calculations and the intermediate results more complicated, with the same final results for the $\zcal$-factors and the soft function. Hence, we choose the simpler regularization scheme here and in the following.

The above bare $cc$ and $\bar{c}\bar{c}$ contributions contain rapidity divergences in the form $1/\nu$, as well as ultraviolet (UV) and infrared (IR) divergences as poles in $\epsilon$. When adding together, the $1/\nu$ singularities cancel, along with logarithms of $\tilde{\mu}$. However, there remains a term proportional to $\ln(-s/m_h^2)$. This give rise to the two-loop soft function $\scal^{(2)}(s,m_h^2)$ defined in Eq.~\eqref{eq:softfunc}. Note that a similar soft function has been given in Ref.~\cite{Becher:2007cu} in QED, with a slight difference: the overall factor $\ln(-s/m_h^2)$ in \eqref{eq:softfunc} is replaced by $\ln(-s/m_Q^2)$ in Eq.~(19) of Ref.~\cite{Becher:2007cu} (converted to our notation). We will see that our convention \eqref{eq:softfunc} is more convenient as it remains the same for massless external legs.

After removing the rapidity divergent parts, we can perform the UV renormalization and extract $\zcal_{[Q]}^{(2),h}$. The UV-renormalized form factor is given by
\begin{equation}
F_1(s,m_Q^2,m_h^2) = Z_Q \left[ F_1^{\text{bare}}(s,m_Q^2,m_h^2) \Big|_{\alpha_s^{\text{bare}} \to Z_{\alpha_s} \alpha_s} \right] ,
\end{equation} 
where $Z_{\alpha_s}$ is the $\overline{\rm MS}$ renormalization constant for the strong coupling $\alpha_s$, and $Z_Q$ is the on-shell field renormalization constant for the massive quark. Neither the one-loop $Z_Q^{(1)}$ nor the one-loop $F_1^{(1),\text{bare}}$ depends on massive quark loops. Hence, for the $n_h$-contributions, we only need the $n_h$-dependent parts of $Z_Q^{(2)}$ and $Z_{\alpha_s}^{(1)}$. We collect these renormalization constants in Appendix~\ref{sec:ren_const}. Finally, we have
\begin{align}\label{eq:Zh2mass}
        \zcal_{[Q]}^{(2),h} &= F_{1,cc}^{(2),\text{bare}}\left(s,m_Q^2,m_h^2\right) + F_{1,\bar{c}\bar{c}}^{(2),\text{bare}}\left(s,m_Q^2,m_h^2\right) \nonumber \\
        & + Z_{\alpha_s}^{(1),h} \left[ F_{1,c}^{(1),\text{bare}}\left(s,m_Q^2\right) + F_{1,\bar{c}}^{(1),\text{bare}}\left(s,m_Q^2\right) \right] + Z_{Q}^{(2)} - C_F \scal^{(2)}(s,m_h^2) \,,
\end{align}
where the one-loop collinear and anti-collinear contributions $F_{1,c/\bar{c}}^{(1),\text{bare}}$ can be found in Ref.~\cite{terHoeve:2023ehm}. Putting everything together, we arrive at the expression in Eq.~\eqref{eq:ZQ2h}.

We now move to the case where both external legs are massless, i.e., $m_1=m_2=0$.
The contributions from $cc$ and $\bar{c}\bar{c}$ regions are given by
\begin{align}
F_{1,cc}^{(2),\text{bare}}\left(s,0,m_h^2\right) &= C_FT_F\left(\frac{\mu^2}{m_h^2}\right)^{2\epsilon} \left( \frac{1}{\nu} + \ln\frac{\tilde{\mu}^2}{-s} \right) \left(\frac{4}{3\epsilon^2} - \frac{20}{9\epsilon} + \frac{112}{27} + \frac{4\zeta_2}{3} \right) \nonumber
\\
&\hspace{-2.5cm} + C_FT_F\left(\frac{\mu^2}{m_h^2}\right)^{2\epsilon}\left[ \frac{4}{3\epsilon^2} - \frac{1}{\epsilon}\left( \frac{8}{9} + \frac{4\zeta_2}{3} \right) + \frac{88}{27} + \frac{32\zeta_2}{9} - \frac{4\zeta_3}{3}\right] \,, \label{eq:F1ccm0}
\\
F_{1,\bar{c}\bar{c}}^{(2),\text{bare}}\left(s,0,m_h^2\right) &= C_FT_F\left(\frac{\mu^2}{m_h^2}\right)^{2\epsilon} \left( \frac{1}{\nu} + \ln\frac{\tilde{\mu}^2}{m_h^2} \right) \left( -\frac{4}{3\epsilon^2} + \frac{20}{9\epsilon} - \frac{112}{27} - \frac{4\zeta_2}{3} \right) \nonumber
\\
&\hspace{-2.5cm}+C_FT_F\left(\frac{\mu^2}{m_h^2}\right)^{2\epsilon}\left[  \frac{2}{\epsilon^3} - \frac{4}{9\epsilon^2} - \frac{1}{\epsilon}\left( \frac{68}{27} - \frac{2\zeta_2}{3} \right) + \frac{124}{9} + \frac{16\zeta_2}{9} - \frac{16\zeta_3}{3}\right]\,.\label{eq:F1cbarcbarm0}
\end{align}
Note that the rapidity divergent parts of Eqs.~\eqref{eq:F1ccm0} and \eqref{eq:F1cbarcbarm0} are the same as those of Eqs.~\eqref{eq:F1cc} and ~\eqref{eq:F1cbarcbar}, respectively. Hence, we may extract the same two-loop soft function $\scal^{(2)}(s,m_h^2)$ here. After that, we can obtain the $\zcal$-factor for massless quarks as
\begin{align}
        \zcal_{[q]}^{(2),h} &= F_{1,cc}^{(2),\text{bare}}\left(s,0,m_h^2\right) + F_{1,\bar{c}\bar{c}}^{(2),\text{bare}}\left(s,0,m_h^2\right) + Z_{q}^{(2)} - C_F\scal^{(2)}(s,m_h^2) \,,
\end{align}
where $Z_{q}^{(2)}$ is the two-loop on-shell wave function renormalization constant for the massless quark $q$. Here we have used the fact that the one-loop collinear and anti-collinear contributions $F_{1,c/\bar{c}}^{(1),\text{bare}}(s,0,m_h^2)$ are scaleless and vanish. The above expression leads to Eq.~\eqref{eq:calZq2}.

The cases where $m_1 \neq m_2$ can be similarly dealt with. In fact, it is easy to see that the double-collinear integrals are independent of $m_2$. In this region, the 5th and 7th propagators in Eq.~\eqref{eq:ff_integral} need to be expanded as
\begin{equation}
-(k_1+k_2+p_2)^2 + m_2^2 \to - \bar{n} \cdot (k_1+k_2) \, n \cdot p_2 \,, \quad (k_1 + p_2)^2 \to \bar{n} \cdot k_1 \, n \cdot p_2 \,.
\end{equation}
The integrals therefore do not involve $p_2^2=m_2^2$. Therefore, the double-collinear contributions simply leads to $F_{1,cc}^{(2),\text{bare}}(s,m_1^2,m_h^2)$. Its expression can be taken from Eq.~\eqref{eq:F1cc} or \eqref{eq:F1ccm0}, depending on whether $m_1$ is zero or not. Similarly, the double-anti-collinear integrals are independent of $m_1^2$ and lead to $F_{1,\bar{c}\bar{c}}^{(2),\text{bare}}(s,m_2^2,m_h^2)$. It is then clear that the same soft function $\scal^{(2)}(s,m_h^2)$ is produced after cancelling the rapidity divergences. The remaining terms coincide with the sum $(\zcal_{[1]}^{(2),h}+\zcal_{[2]}^{(2),h})/2$.

\begin{figure}[t!]
	\centering
	\subfigure{
        \includegraphics[viewport = 36 258 576 756,width=0.17\textwidth]{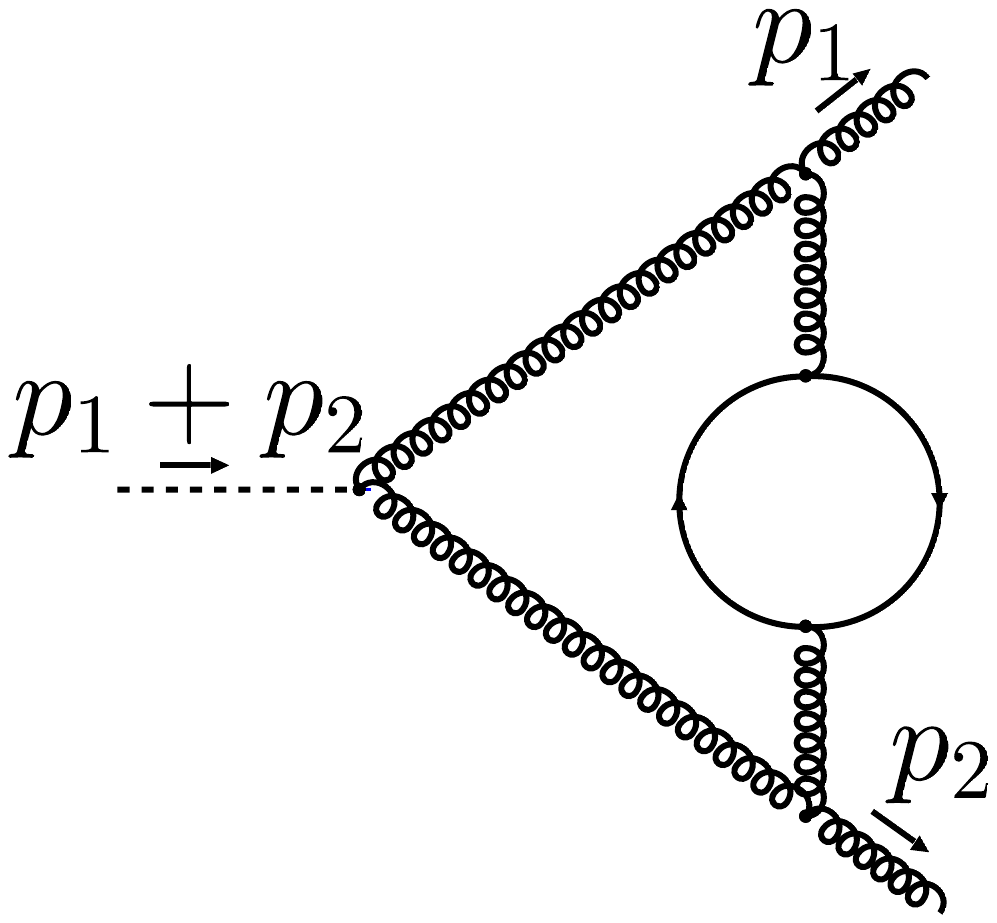}
        }
        \subfigure{
        \includegraphics[viewport = 36 315 576 756,width=0.17\textwidth]{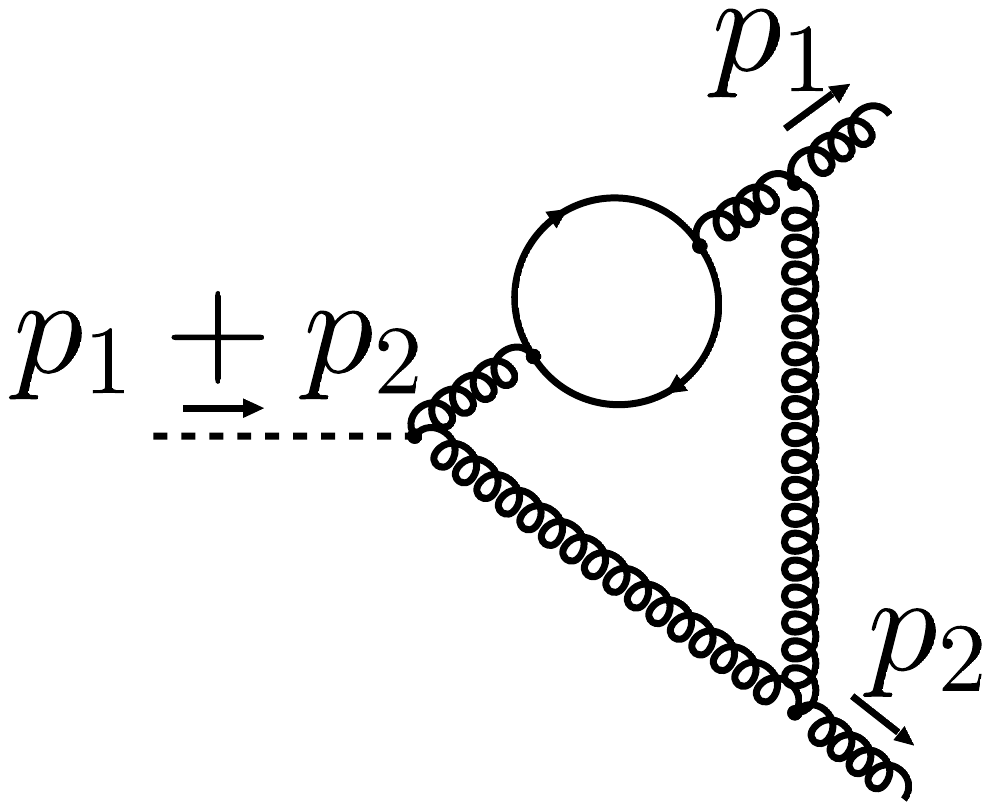}
        }
        \subfigure{
        \includegraphics[viewport = 36 315 576 756,width=0.17\textwidth]{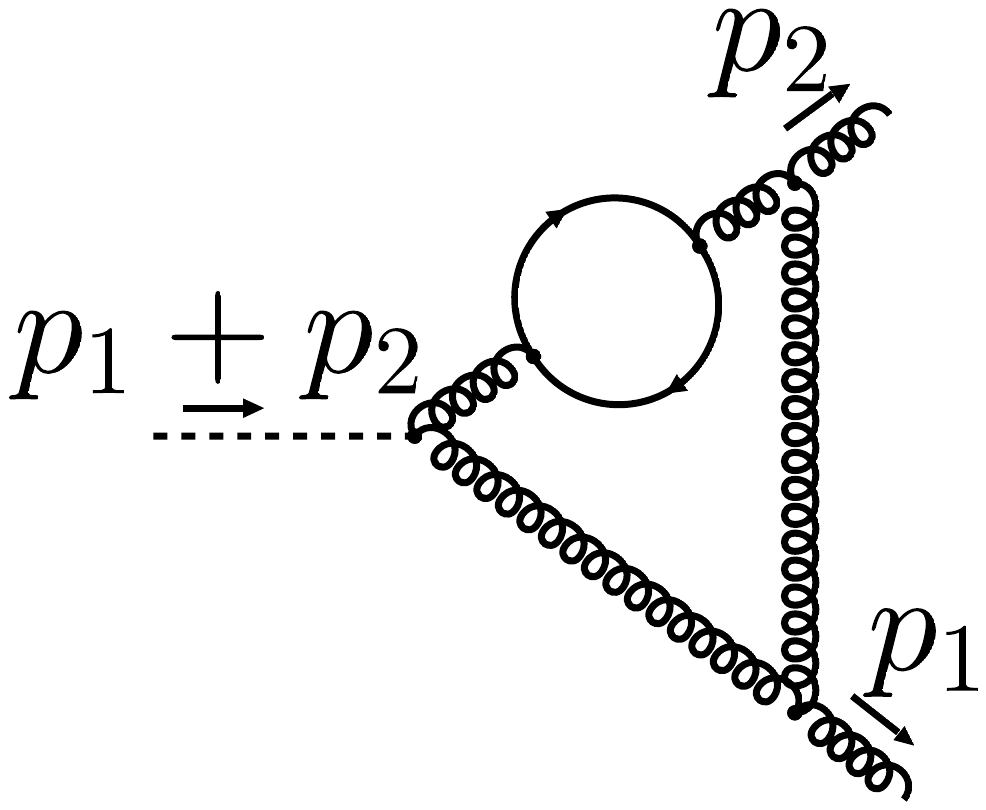}
        }
        \subfigure{
        \includegraphics[viewport = 36 311 576 756,width=0.17\textwidth]{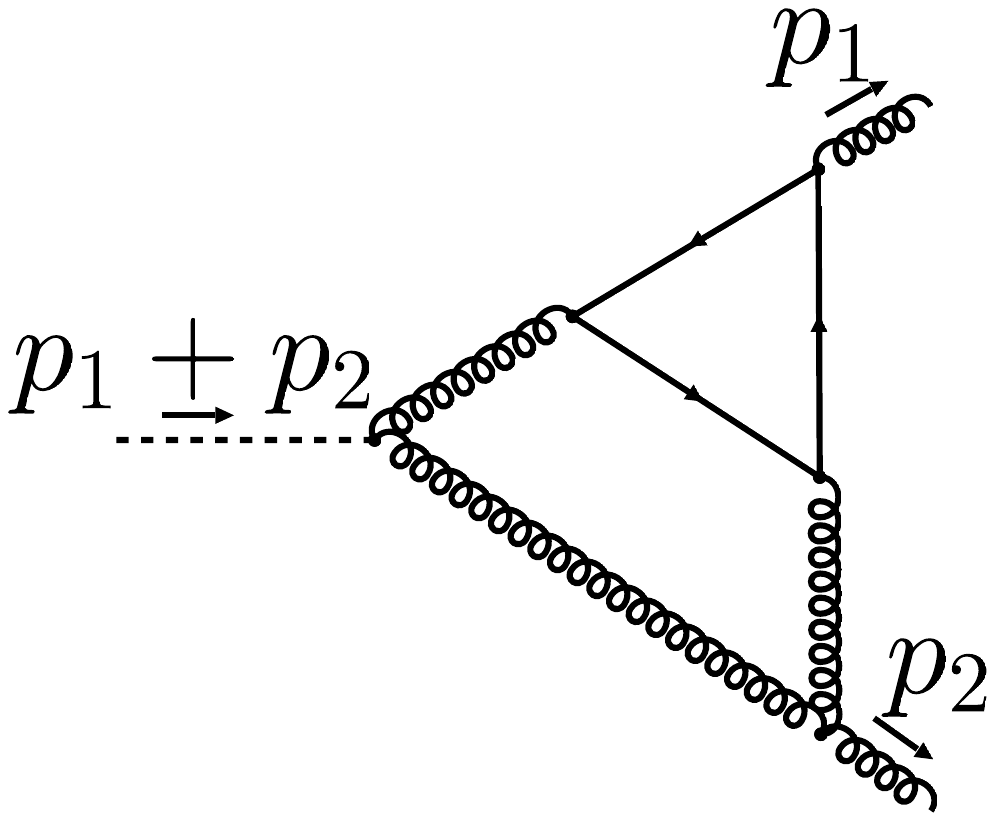}
        }
        \subfigure{
        \includegraphics[viewport = 36 311 576 756,width=0.17\textwidth]{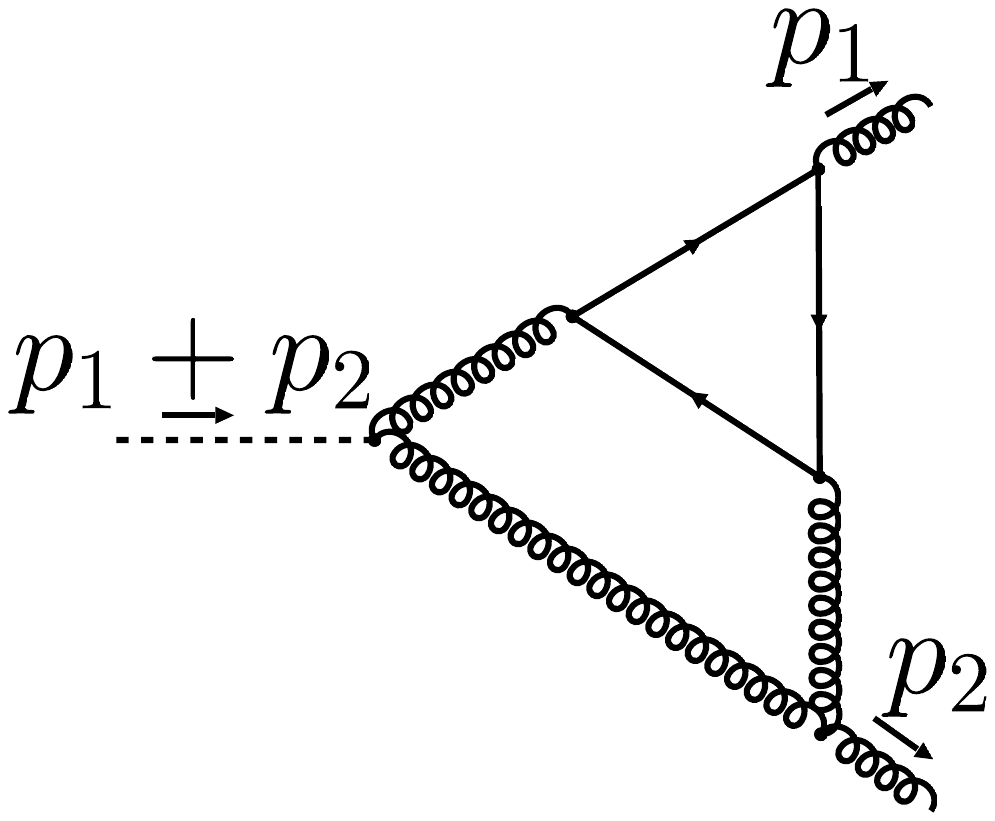}
        }
        \subfigure{
        \includegraphics[viewport = 36 311 576 756,width=0.17\textwidth]{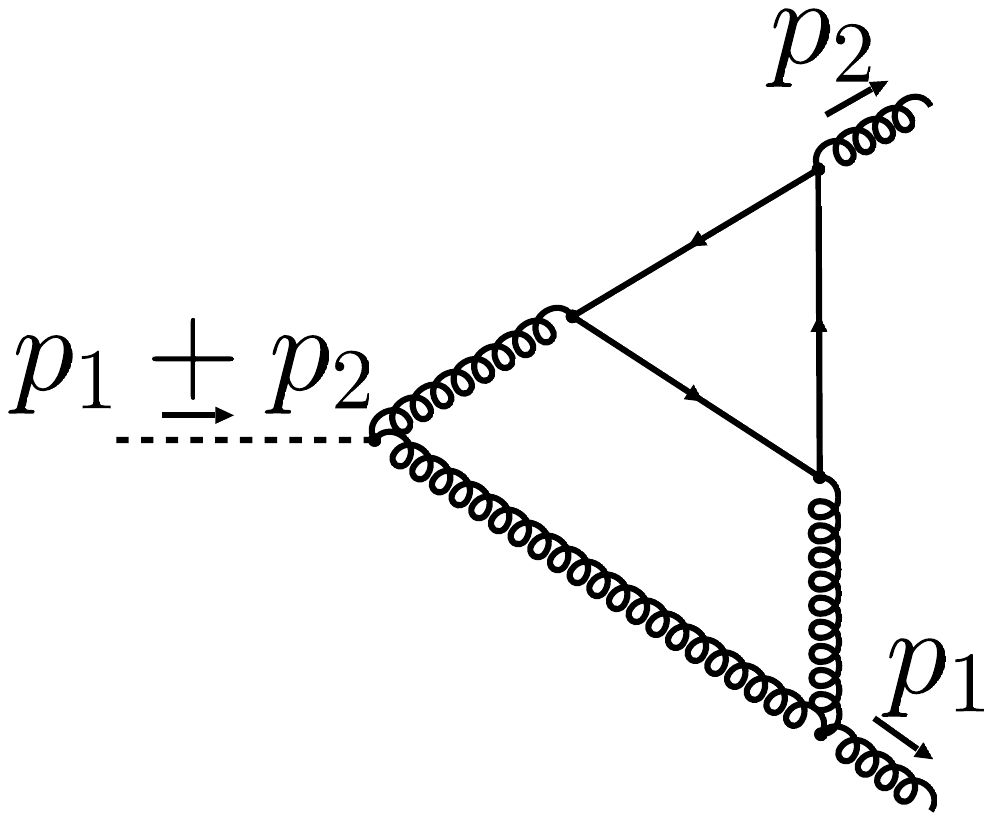}
        }
        \subfigure{
        \includegraphics[viewport = 36 311 576 756,width=0.17\textwidth]{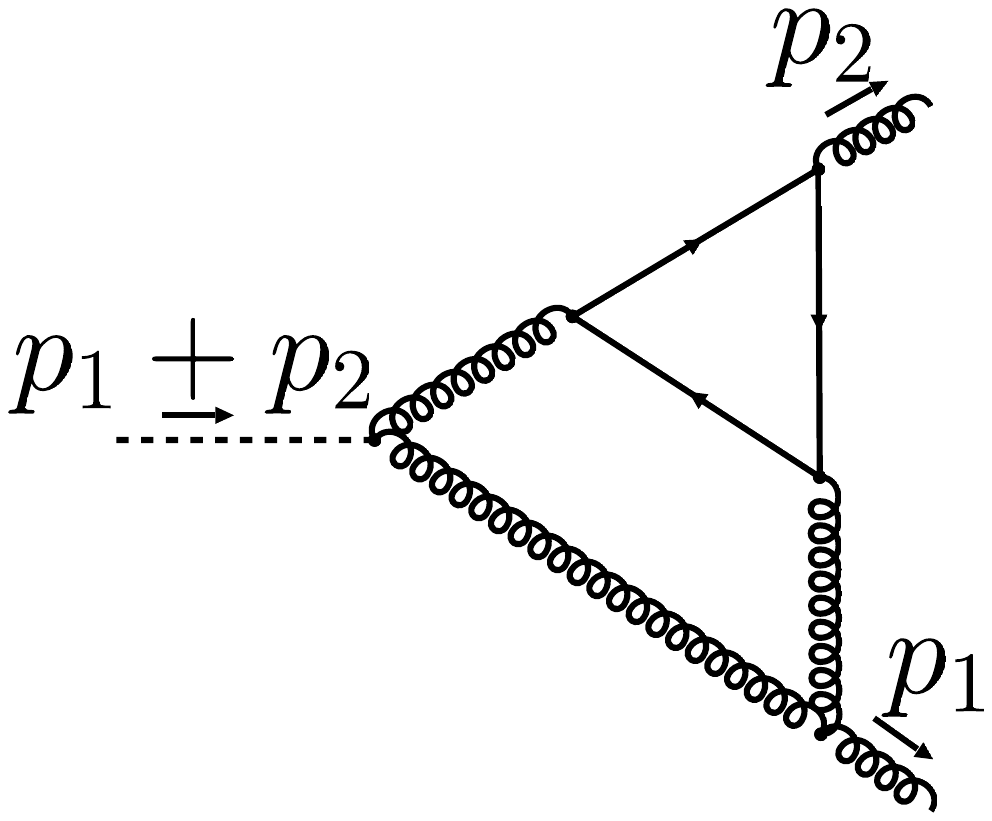}
        }
        \subfigure{
        \includegraphics[viewport = 36 310 576 756,width=0.17\textwidth]{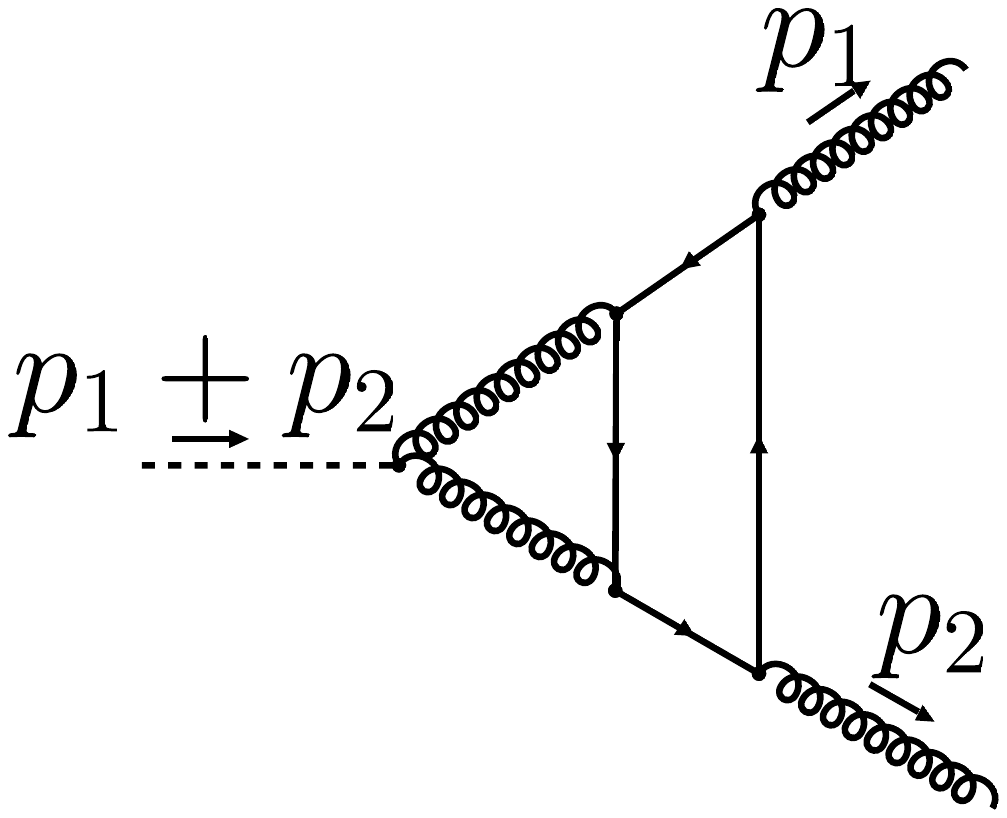}
        }
        \subfigure{
        \includegraphics[viewport = 36 310 576 756,width=0.17\textwidth]{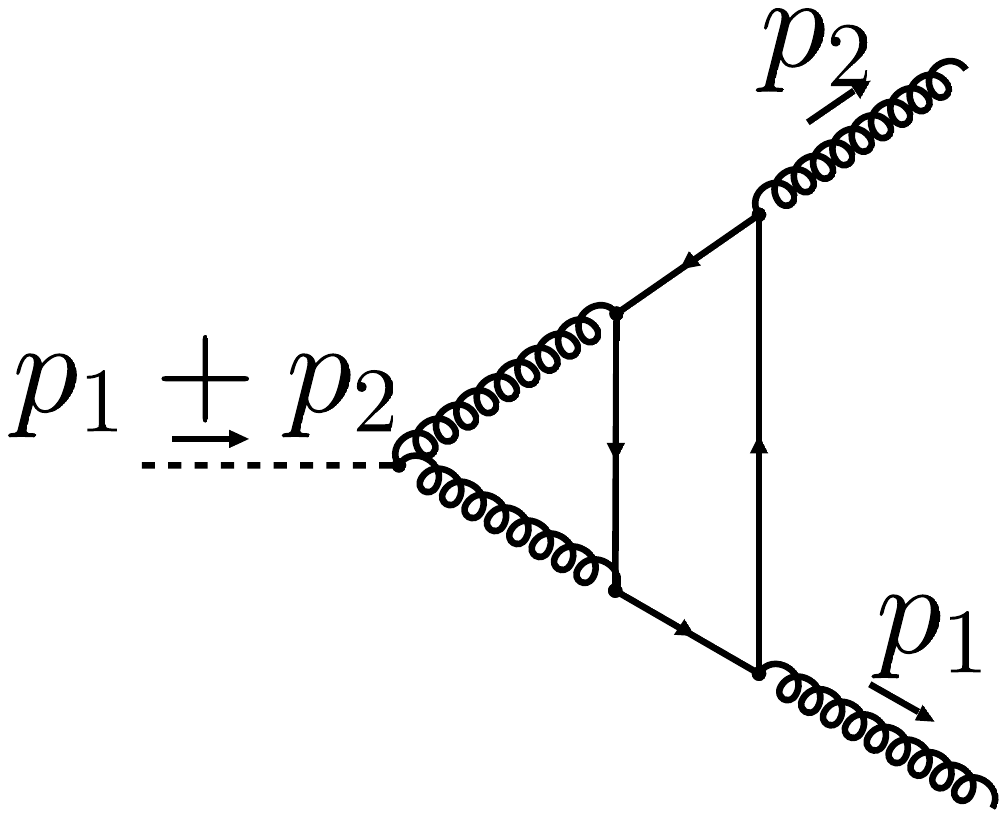}
        }
        \subfigure{
        \includegraphics[viewport = 36 312 576 756,width=0.17\textwidth]{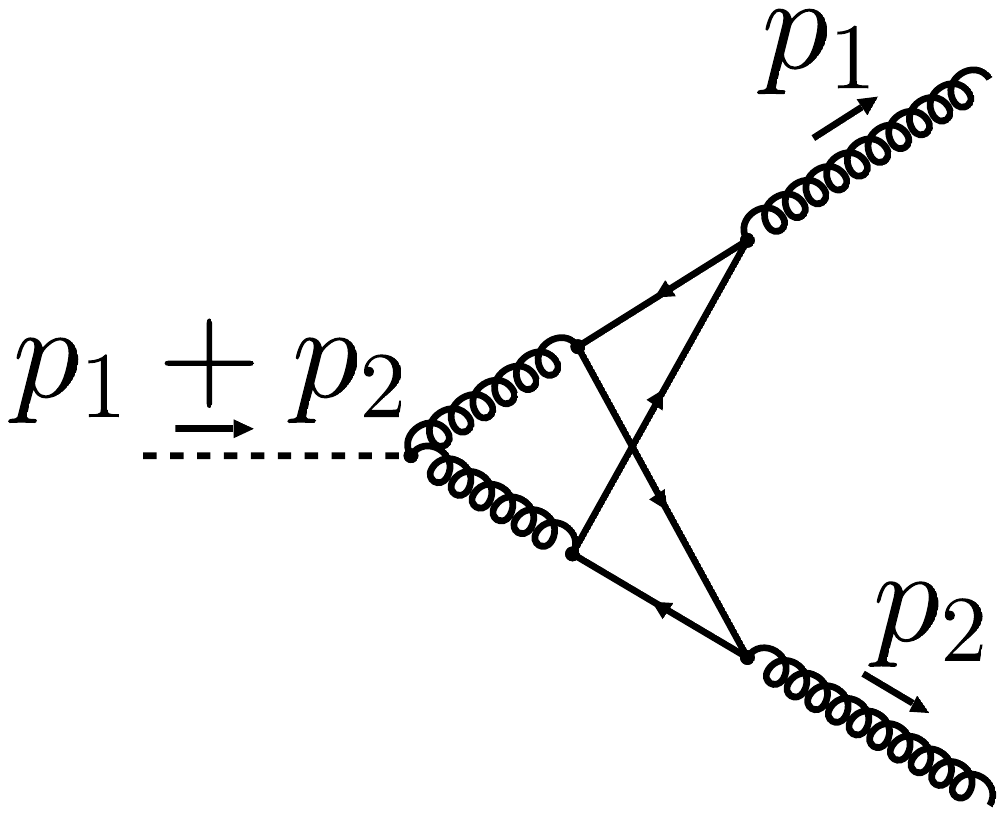}
        }
        \subfigure{
        \includegraphics[viewport = 36 470 576 756,width=0.17\textwidth]{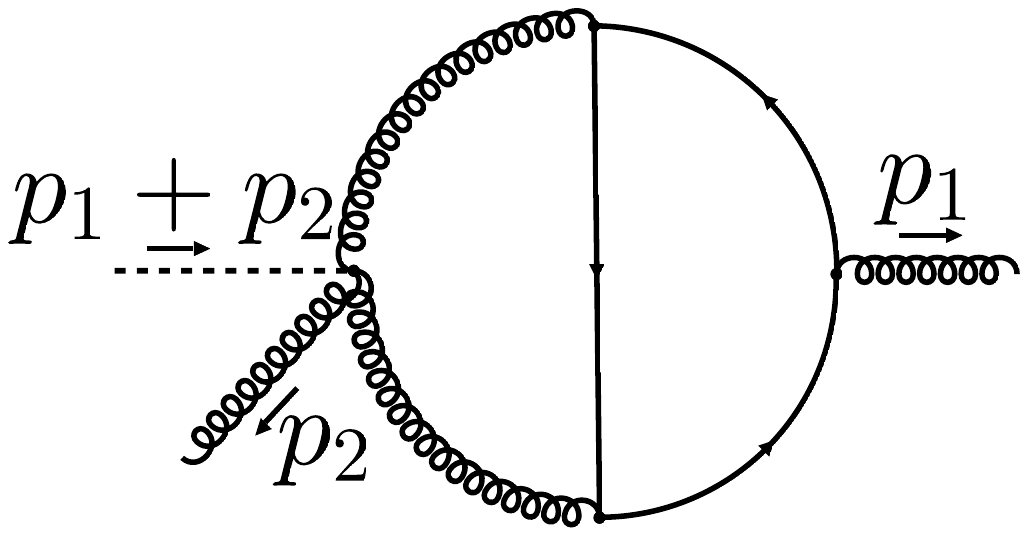}
        }
        \subfigure{
        \includegraphics[viewport = 36 470 576 756,width=0.17\textwidth]{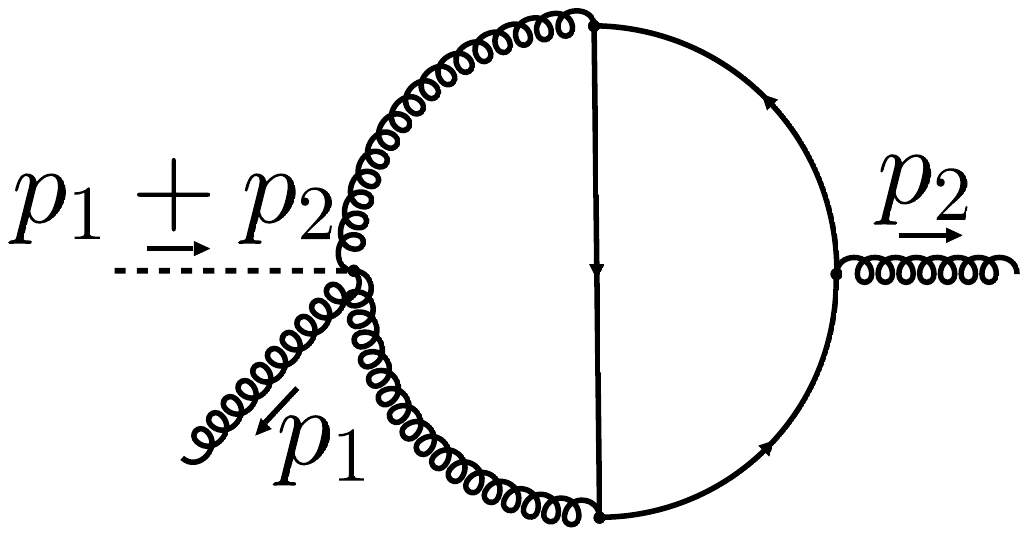}
        }
        \subfigure{
        \includegraphics[viewport = 36 450 576 756,width=0.17\textwidth]{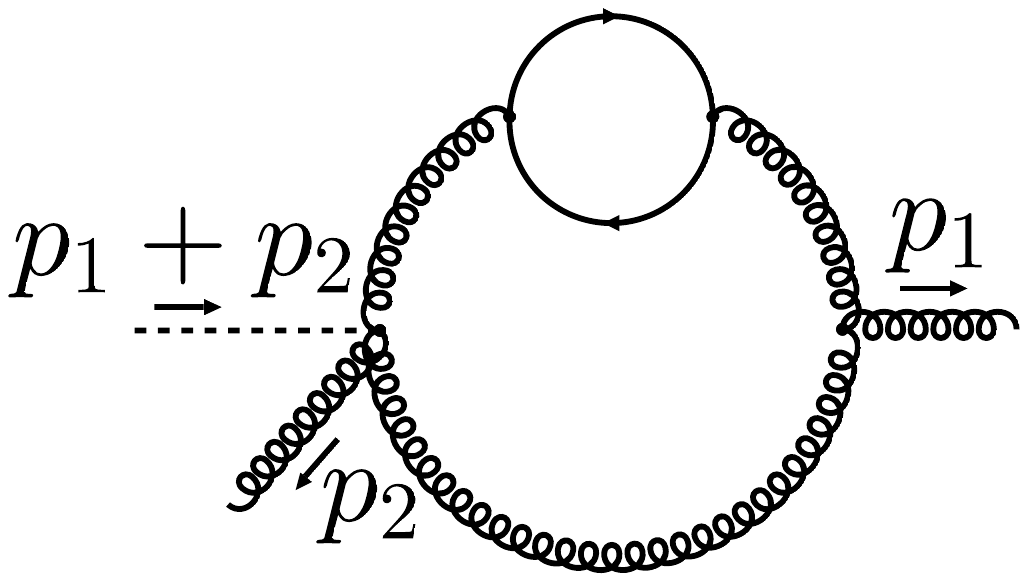}
        }
        \subfigure{
        \includegraphics[viewport = 36 450 576 756,width=0.17\textwidth]{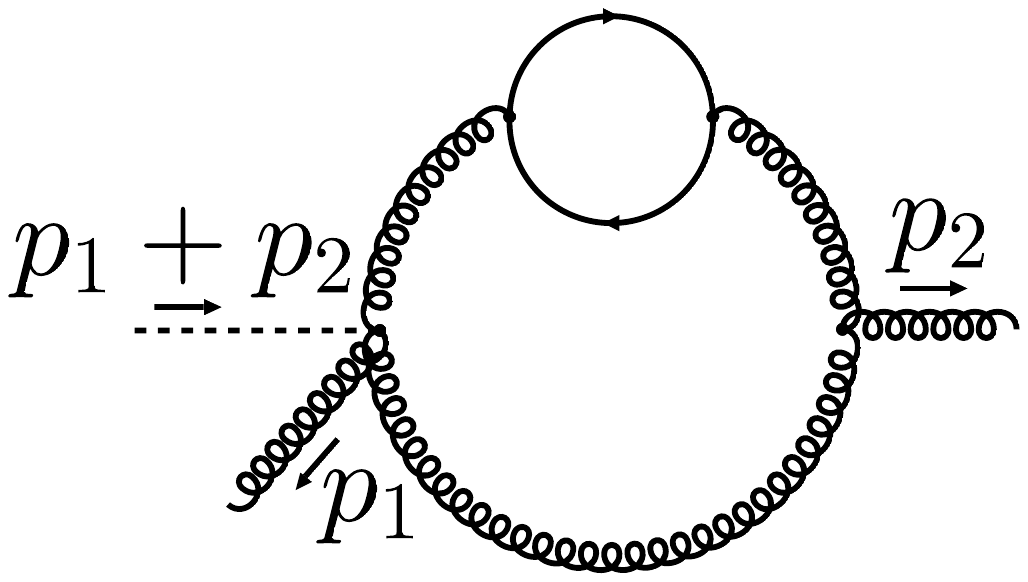}
        }
        \subfigure{
        \includegraphics[viewport = 36 483 576 756,width=0.17\textwidth]{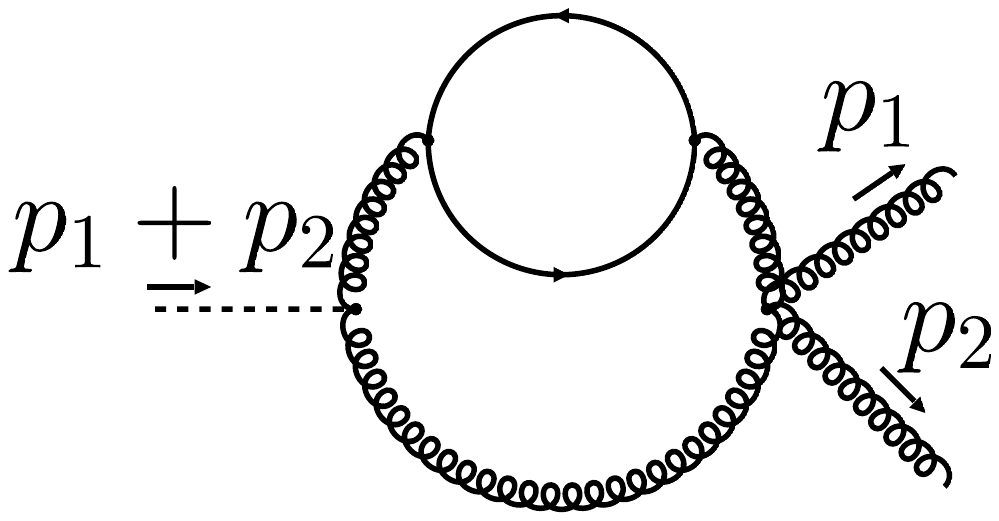}
        }
	\caption{Two-loop Feynman diagrams involving a massive quark loop for the gluon form factor.}
	\label{fig:diagramg}
\end{figure}

Finally, we apply the same methodology to the gluon case and derive $\zcal_{[g]}^{(2),h}$. We start from the gluon scalar form factor generated from the effective Lagrangian
\begin{align}
    {\cal L}_{\text{int}}=-\frac{\lambda}{4}H G_a^{\mu \nu } G_{a,\mu \nu } \,.
\end{align}
From the effective $H$-$g(p_1)$-$g(p_2)$ vertex $\Gamma^{\mu \nu}_{gg}$, we project out the scalar form factor as
\begin{equation}
        {F}_{gg} = \frac{p_1\cdot p_2 \, g_{\mu\nu} - p_{1,\mu} p_{2,\nu} - p_{1,\nu} p_{2,\mu} }{2 (1-{\epsilon})} \, \Gamma_{gg}^{\mu \nu} \, . \label{eq:projg}
\end{equation}
We use \texttt{QGRAF}~\cite{Nogueira:1991ex} to generate the relevant two-loop Feynman diagrams involving a massive quark loop, shown in Fig.~\ref{fig:diagramg}. We then use \texttt{FORM} \cite{Vermaseren:2000nd} to manipulate the resulting expressions. In the gluon cases, there are quark-box and quark-triangle insertions besides the bubble insertions. Using the method of regions, we find that, in the high-energy limit at LP, the diagrams with a quark-box insertion do not contribute in the $cc$ and $\bar{c}\bar{c}$ regions. To determine $\zcal_{[g]}^{(2),h}$, we therefore only need to consider the contributions from the quark-triangle and bubble insertions. The scalar integrals appearing in these diagrams can be expressed in the following generic form
\begin{multline}
\label{eq:ff_integral_gg}
I^A_{\{a_i\}}
\equiv \mu^{4\epsilon}\int\frac{dk_1}{(2\pi)^d}\frac{dk_2}{(2\pi)^d}
\frac{1}{[k_1^2-m_h^2]^{a_1}}
\frac{1}{[k_2^2-m_h^2]^{a_2}}
\frac{1}{[(k_1+k_2)^2]^{a_3}}
\frac{1}{[(k_1+k_2-p_1)^2]^{a_4}}
\\
\times
\frac{\left(-\tilde{\mu}^2\right)^{\nu}}{[(k_1+k_2+p_2)^2]^{a_5+\nu}}
\frac{1}{[(k_1-p_1)^2-m_h^2]^{a_6}}
\frac{1}{[(k_1+p_2)^2-m_h^2]^{a_7}}\,.
\end{multline}
Note that only the first diagram in Fig.~\ref{fig:diagramg} contains rapidity divergences. Hence, we may set $\nu \to 0$ from the beginning for the other diagrams. Only the $cc$ and $\bar{c}\bar{c}$ regions are relevant for $\zcal_{[g]}^{(2),h}$, and their contributions are given by
\begin{align}
F_{gg,cc}^{(2),\text{bare}}\left(s,m_h^2\right) &= C_AT_F\left(\frac{\mu^2}{m_h^2}\right)^{2\epsilon} \left( \frac{1}{\nu} + \ln\frac{\tilde{\mu}^2}{-s} \right) \left(\frac{4}{3\epsilon^2} - \frac{20}{9\epsilon} + \frac{112}{27} + \frac{4\zeta_2}{3} \right) \nonumber
\\
&\hspace{-2.5cm} + C_AT_F\left(\frac{\mu^2}{m_h^2}\right)^{2\epsilon}\left[ \frac{5}{6\epsilon^2} - \frac{11}{36\epsilon} + \frac{179}{216} + \frac{5\zeta_2}{6} \right] , \label{eq:Fggccm0}
\\
F_{gg,\bar{c}\bar{c}}^{(2),\text{bare}}\left(s,m_h^2\right) &= C_AT_F\left(\frac{\mu^2}{m_h^2}\right)^{2\epsilon} \left( \frac{1}{\nu} + \ln\frac{\tilde{\mu}^2}{m_h^2} \right) \left( -\frac{4}{3\epsilon^2} + \frac{20}{9\epsilon} - \frac{112}{27} - \frac{4\zeta_2}{3} \right) \nonumber
\\
&\hspace{-2.5cm}+C_AT_F\left(\frac{\mu^2}{m_h^2}\right)^{2\epsilon}\left[  \frac{2}{\epsilon^3} - \frac{17}{18\epsilon^2} - \frac{1}{\epsilon}\left( \frac{209}{108} - 2\zeta_2 \right) + \frac{817}{72} - \frac{17\zeta_2}{18} - 4\zeta_3\right]\,.\label{eq:Fggcbarcbarm0}
\end{align}
Note that the rapidity divergent parts of Eqs.~\eqref{eq:Fggccm0} and \eqref{eq:Fggcbarcbarm0} are the same as those of Eqs.~\eqref{eq:F1cc} and ~\eqref{eq:F1cbarcbar}, respectively, with the color factor $C_F \to C_A$. We may then extract the same two-loop soft function $\scal^{(2)}(s,m_h^2)$ here. After that, we extract the  $\zcal$-factor for gluon according to
\begin{align}
        \zcal_{[g]}^{(2),h} &= F_{gg,cc}^{(2),\text{bare}}\left(s,m_h^2\right) + F_{gg,\bar{c}\bar{c}}^{(2),\text{bare}}\left(s,m_h^2\right) + Z_{g}^{(2)} - C_A\scal^{(2)}(s,m_h^2) \,,
\end{align}
where $Z_{g}^{(2)}$ is the two-loop on-shell renormalization constant for gluon fields. The above expression leads to Eq.~\eqref{eq:calZg2}.

\subsection{The quark form factors}

We now turn to the validation of our factorization formula \eqref{eq:fac} in the context of quark form factors. We will only be concerned with the $n_h$-dependent part, since it is the new result of this work. On the right-hand side of \eqref{eq:fac}, we need the two-loop massless quark form factor with a massless quark-bubble insertion. This can be extracted from \cite{Moch:2005id, Gehrmann:2005pd}, and is given by
\begin{align}
F_{1,q\bar{q}}^{(2),l}(s) &= C_F T_F \biggl[ -\frac{2}{\epsilon^3} + \frac{1}{\epsilon^2} \left( \frac{4L_s}{3} - \frac{8}{9} \right) + \frac{1}{\epsilon} \left( \frac{65}{27} - \frac{20L_s}{9} + 2\zeta_2 \right) \nonumber
\\
&- \frac{4L_s^3}{9} + \frac{38L_s^2}{9} - \frac{418L_s}{27} - \frac{8\zeta_2L_s}{3} + \frac{4085}{162} + \frac{46\zeta_2}{9} + \frac{4\zeta_3}{9} \biggr] \,,
\end{align}
where $L_s = \ln(-s/\mu^2)$.

The massive-massive quark form factor with a massive quark-bubble insertion in the special case $m_1 = m_2 = m_h$ can be extracted from \cite{Bernreuther:2004ih, Gluza:2009yy, Ablinger:2017hst}, where $m_1$ and $m_2$ denote the masses of two external quarks, and $m_h$ is the mass of the internal bubble. The result reads
\begin{align}
F_{1,Q\bar{Q}}^{(2),h}(s,m_h^2,m_h^2,m_h^2) &= C_F T_F \biggl[ \frac{8}{3\epsilon} \left( L_m - L_m^2 - L_m L_s \right) - \frac{28L_m^3}{9} - \frac{4L_s^3}{9} - \frac{8L_m^2L_s}{3} \nonumber
\\
&+ \frac{38L_m^2}{9} + \frac{38L_s^2}{9} + \frac{40L_mL_s}{9} - \frac{20\zeta_2L_m}{3} -4\zeta_2L_s \nonumber
\\
&- \frac{386L_m}{27} - \frac{530L_s}{27} + \frac{1532}{27} - \frac{8\zeta_2}{3} \biggr] \,,
\end{align}
where $L_m = \ln(\mu^2/m_h^2)$. It is straightforward to check that it satisfies the factorization formula:
\begin{equation}
F_{1,Q\bar{Q}}^{(2),h}(s,m_h^2,m_h^2,m_h^2) = F_{1,q\bar{q}}^{(2),l}(s) + C_F \scal^{(2)}(s,m_h^2) + \zcal_{[Q]}^{(2),h}(m_h^2,m_h^2) \,.
\end{equation}

We now turn to the massive-massless quark form factor with a massive quark-bubble insertion, in the special case $m_1=m_h$. This can be extracted from \cite{Bonciani:2008wf, Asatrian:2008uk, Beneke:2008ei, Bell:2008ws}, and the expression is
\begin{align}
F_{Q\bar{q}}^{(2),h}(s,m_h^2,m_h^2) &= C_F T_F \biggl[ \frac{4L_m}{3\epsilon^2} + \frac{1}{\epsilon} \left( \frac{10L_m}{3} - \frac{2L_m^2}{3} - \frac{8L_mL_s}{3} + \frac{2\zeta_2}{3} \right) \nonumber
\\
&- \frac{14L_m^3}{9} - \frac{4L_s^3}{9} - \frac{8L_m^2L_s}{3} + \frac{47L_m^2}{9} + \frac{38L_s^2}{9} + \frac{40L_mL_s}{9} \nonumber
\\
&- \frac{14\zeta_2L_m}{3} - 4\zeta_2 L_s - \frac{314L_m}{27} - \frac{530L_s}{27}  + \frac{7951}{162} + \frac{35\zeta_2}{9} - \frac{28\zeta_3}{9} \biggr] \,.
\end{align}
The corresponding factorization formula can be verified to hold:
\begin{equation}
F_{Q\bar{q}}^{(2),h}(s,m_h^2,m_h^2) = F_{q\bar{q}}^{(2),l}(s) + C_F \scal^{(2)}(s,m_h^2) + \frac{1}{2} \zcal_{[Q]}^{(2),h}(m_h^2,m_h^2) + \frac{1}{2} \zcal_{[q]}^{(2),h}(m_h^2) \,.
\end{equation}

Finally, we consider the two-loop massless-massless quark form factor with a massive-quark bubble insertion. This is not available in the literature, so we have to calculate that ourselves. There is only one diagram and it is straightforward to write down the loop integrands. For the loop integrals, we refer to the next sub-section, since they will also appear in the gluon-gluon form factors. The unrenormalized two-loop form factor is then given by
\begin{align}
        F_{q\bar{q}}^{(2),h,\text{bare}} &= C_F T_F \biggl[\frac{8}{3 \epsilon ^3}+\frac{1}{\epsilon ^2}\left(\frac{8 L_m}{3}-\frac{8 L_s}{3}+4\right)\nonumber \\
                     &\qquad\qquad+\frac{1}{\epsilon }\left(-\frac{8 L_m L_s}{3}+\frac{4 L_m^2}{3}+4 L_m+\frac{4 L_s^2}{3}-4 L_s+\frac{29}{3}\right)\nonumber \\
                     &\qquad\qquad-\frac{8}{3} L_m^2 L_s+\frac{40 L_m L_s}{9}+\frac{56 L_m^2}{9}-\frac{8 \zeta _2 L_m}{3}-\frac{296 L_m}{27}-\frac{8 L_s^3}{9}\nonumber \\
                     &\qquad\qquad+\frac{56 L_s^2}{9} -\frac{8 \zeta _2 L_s}{3}-\frac{818 L_s}{27}-\frac{112 \zeta_3}{9}+\frac{76 \zeta _2}{9}+\frac{10301}{162}\biggl] \,.
\end{align}

After renormalization, we get
\begin{align}
    F_{q\bar{q}}^{(2),h}(s,m_h^2) &= C_F T_F \biggl[ \frac{8 L_m}{3 \epsilon ^2}+\frac{1}{\epsilon}\left(-\frac{8 L_m L_s}{3}+\frac{4 L_m^2}{3}+4 L_m+\frac{4 \zeta _2}{3}\right)\nonumber \\
              &-\frac{56 \zeta _3}{9}-\frac{8}{3} L_m^2 L_s+\frac{40 L_m L_s}{9}+\frac{56 L_m^2}{9}-\frac{8 \zeta _2 L_m}{3}\nonumber \\
              &-\frac{242 L_m}{27}-\frac{4 L_s^3}{9}+\frac{38 L_s^2}{9}-4 \zeta _2 L_s-\frac{530 L_s}{27}+\frac{94 \zeta _2}{9}+\frac{3355}{81}\biggl] \,.
\end{align}

Again, we find that it satisfies the factorization formula:
\begin{equation}
F_{q\bar{q}}^{(2),h}(s,m_h^2) = F_{q\bar{q}}^{(2),l}(s) + C_F \scal^{(2)}(s,m_h^2) + \zcal_{[q]}^{(2),h}(m_h^2) \,.
\end{equation}

\subsection{The gluon form factor}


For the gluon form factor, we again consider the diagrams in Fig.~\ref{fig:diagramg}. We now need to calculate the integrals without region expansion. The integrals can be categorized into two different families. The first family (family A) covers all planar diagrams, and is defined in Eq.~\eqref{eq:ff_integral_gg}. These integrals can be expressed in terms of multiple polylogarithms (MPLs), and can be found in \cite{Bonciani:2003hc, Anastasiou:2006hc, Bonciani:2003te, Davies:2018ood}. The second family (family B) corresponds to the non-planar diagram (the last one in the second row of Fig.~\ref{fig:diagramg}), and is defined as
\begin{multline}
\label{eq:ff_integral_gg_non-planar}
I^B_{\{a_i\}}
\equiv \mu^{4\epsilon}\int\frac{dk_1}{(2\pi)^d}\frac{dk_2}{(2\pi)^d}
\frac{1}{[k_1^2]^{a_1}}
\frac{1}{[(k_1-p_1-p_2)^2]^{a_2}}
\frac{1}{[k_2^2-m_h^2]^{a_3}}
\frac{1}{[(k_2+p_1)^2-m_h^2]^{a_4}}
\\
\times
\frac{1}{[(k_1+k_2)^2-m_h^2]^{a_5}}
\frac{1}{[(k_1+k_2-p_2)^2-m_h^2]^{a_6}}
\frac{1}{[(k_2+p_2)^2]^{a_7}} \,.
\end{multline} 
This family involves elliptic integrals. In the high-energy limit $|s| \gg m_h^2$, the asymptotic expressions for integrals in family B can be found in \cite{Jiang:2023jmk}. For the UV renormalization, we need $Z_g$ for the gluon field, $Z_{\alpha_s}$ for the strong coupling, as well as $Z_\lambda$ for the effective operator. These renormalization constants are collected in Appendix~\ref{sec:ren_const}. After renormalization, the two-loop gluon form factor with a massive quark loop is given by

\begin{align}
    F_{gg}^{(2),h}(s,m_h^2) = &\frac{1}{N_c}T_F \left(-8 \zeta _3+2L_m-2L_s+\frac{56}{3}\right)\nonumber \\
                & +N_cT_F \biggl[ \frac{16 L_m}{3 \epsilon ^2} + \frac{1}{\epsilon }\left(-\frac{16 L_m L_s}{3}+\frac{8 L_m^2}{3}+\frac{88 L_m}{9}+\frac{8 \zeta _2}{3}\right)\nonumber \\
                & \qquad-\frac{64 \zeta _3}{9}-4 L_m^2 L_s+\frac{4}{3} L_m L_s^2+\frac{40 L_m L_s}{9}+\frac{4 L_m^3}{9}+\frac{64 L_m^2}{9}\nonumber \\
                & \qquad+\frac{8 \zeta _2 L_m}{3}-\frac{422 L_m}{27} -\frac{4 L_s^3}{9}+\frac{20 L_s^2}{9}+\frac{46 L_s}{27}+\frac{4 \zeta _2}{9}-\frac{2270}{81}\biggl] \,,
\end{align}
where $N_c$ is the number of colors. Here, it is worth mentioning that the massive quark-loop contributions to $F_{gg}^{(2)}$ consist of three parts: the contribution from a single massive quark loop as shown above, the contribution from two massive quark loops (which we denote as the $n_h^2$ contribution), and the contribution from a massive quark loop and a massless quark loop (which we denote as the $n_h n_l$ contribution). The last two parts arise from renormalization, and are not included in the above expression. In the factorization formula, the $n_h^2$ and $n_h n_l$ contributions are taken care of by the $(\zcal_{[g]}^{(1)})^2$ term and the $(n_l+n_h)\zcal_{[g]}^{(1)}$ term in Eq.~\eqref{eq:zcal_g_2}, which were already known in \cite{Mitov:2006xs}.

For the purpose of this work, we only need to check the factorization property of $F_{gg}^{(2),h}$. For that we need the one-loop gluon form factor without quark loops, and the two-loop gluon form factor with a massless quark loop. They can be extracted from \cite{Gehrmann:2005pd, Gehrmann:2010ue} and are given by
\begin{align}
        F_{gg}^{(1),\text{no-quark}} &= N_cT_F \biggl[-\frac{4}{\epsilon ^2}+\frac{1}{\epsilon}\left(4 L_s-\frac{22}{3}\right)+ 2\zeta _2-2L_s^2\nonumber \\
                  &\qquad+\epsilon\left(\frac{28 \zeta _3}{3}+\frac{2L_s^3}{3}-2\zeta _2 L_s-4\right)\nonumber \\
		  &\qquad+\epsilon^2\left(-\frac{28 \zeta _3 L_s}{3}-\frac{L_s^4}{6}+ \zeta _2 L_s^2+4 L_s+\frac{47 \zeta _4}{4}-12\right)\biggl] \,,
\\
        F_{gg}^{(2),l} &= \frac{1}{N_c}T_F \left( -\frac{1}{\epsilon }-8 \zeta _3-2L_s+\frac{67}{6}\right)\nonumber \\
                  &+N_c \biggl[-\frac{14}{3 \epsilon ^3}+\frac{1}{\epsilon ^2}\left(4 L_s-\frac{26}{3}\right)+\frac{1}{\epsilon}\left(-\frac{4 L_s^2}{3}-\frac{20 L_s}{9}+\frac{2\zeta _2}{3}+\frac{155}{27}\right)\nonumber \\
                  &\qquad\qquad+4 \zeta _3+\frac{20 L_s^2}{9}+\frac{4 \zeta _2 L_s}{3}+\frac{158 L_s}{27}-\frac{10 \zeta _2}{3}-\frac{5905}{162}\biggl] \,.
\end{align}
It is then straightforward to check the factorization formula:
\begin{equation}
        F_{gg}^{(2),h}(s,m_h^2) = F_{gg}^{(2),l}(s) +\zcal_{[g]}^{(1)}(m_h^2) \, F_{gg}^{(1),\text{no-quark}}(s) + \zcal_{[g]}^{(2),h}(m_h^2) + N_c \, \scal^{(2)}(s,m_h^2) \,.
\end{equation}

\subsection{The scattering amplitudes for \texorpdfstring{$t\bar{t}$}{} production}

As the last and the most complicated validation, we consider the $q\bar{q} \to t\bar{t}$ and $gg \to t\bar{t}$ partonic processes. The two-loop amplitudes in the high-energy limit have been calculated in \cite{Czakon:2007ej, Czakon:2007wk}. The purely massless amplitudes for $q\bar{q} \to q'\bar{q}'$ and $gg \to q'\bar{q}'$ are given in \cite{Anastasiou:2000kg, Anastasiou:2001sv}. In contrast to the form factor cases where the color structure is simple, the soft function in $t\bar{t}$ production is a non-trivial operator in color space. To deal with this, we follow the same strategy of \cite{Ferroglia:2009ii}. For the $q_l + \bar{q}_k \to t_i + \bar{t}_j$ channel we choose the color basis as
\begin{equation}
\ket{c_1} = \delta_{ij} \delta_{kl} \,, \quad \ket{c_2} = \left( t^a \right)_{ij} \left( t^a \right)_{kl} \,,
\end{equation}
and for the $g_a + g_b \to t_i + \bar{t}_j$ channel we use
\begin{equation}
\ket{c_1} = \delta^{ab} \delta_{ij} \,, \quad \ket{c_2} = i f^{abc} \left( t^c \right)_{ij} \,, \quad \ket{c_3} = d^{abc} \left( t^c \right)_{ij} \,,
\end{equation}
where $i,j,k,l,a,b$ are color indices. The scattering amplitudes are then projected into the above bases:
\begin{equation}
\Ket{\mathcal{M}} = \sum_I \mathcal{M}_I \ket{c_I} \,.
\end{equation}
Using this decomposition, we may rewrite Eq.~\eqref{eq:fac} as
\begin{equation}
\mathcal{M}^{\text{massive}}_I = \sum_J \prod_i \left(\zcal_{[i]}^{(m|0)}\right)^{1/2}
\bm{\scal}_{IJ} \, \mathcal{M}^{\text{massless}}_J \, ,
\end{equation}
where the soft matrix elements are defined by
\begin{equation}
\bm{\scal}_{IJ} = \frac{\braket{c_I | \bm{\scal} | c_J}}{\braket{c_I | c_I}}  \,.
\end{equation}
For the $q\bar{q}$ channel, the matrix elements of $\bm{T}_i \cdot \bm{T}_j$ in $\bm{\scal}$ (see Eq.~\eqref{eq:soft_operator}) are given by
\begin{align}
\bm{T}_1 \cdot \bm{T}_2 = \bm{T}_3 \cdot \bm{T}_4 &=
\begin{pmatrix}
-C_F & 0
\vspace{1mm}\\
0 & \frac{1}{2N_c}
\end{pmatrix}
, \nonumber
\\
\bm{T}_1 \cdot \bm{T}_3 = \bm{T}_2 \cdot \bm{T}_4  &=
\begin{pmatrix}
0 & -\frac{C_F}{2N_c}
\vspace{1mm}\\
-1 & -\frac{N_c^2-2}{2N_c}
\end{pmatrix}
, \nonumber
\\
\bm{T}_2 \cdot \bm{T}_3 = \bm{T}_1 \cdot \bm{T}_4 &=
\begin{pmatrix}
0 & \frac{C_F}{2N_c}
\vspace{1mm}\\
1 & -\frac{1}{N_c}
\end{pmatrix}
.
\end{align}
For the $gg$ channel, they are
\begin{align}
\bm{T}_1 \cdot \bm{T}_2 &=
\begin{pmatrix}
-N_c & 0 & 0
\vspace{1mm}\\
0 & -\frac{N_c}{2} & 0
\vspace{1mm}\\
0 & 0 & -\frac{N_c}{2}
\end{pmatrix}
, \nonumber
\\
\bm{T}_3 \cdot \bm{T}_4 &=
\begin{pmatrix}
-C_F & 0 & 0
\vspace{1mm}\\
0 & \frac{1}{2N_c} & 0
\vspace{1mm}\\
0 & 0 & \frac{1}{2N_c}
\end{pmatrix}
, \nonumber
\\
\bm{T}_1 \cdot \bm{T}_3 = \bm{T}_2 \cdot \bm{T}_4  &=
\begin{pmatrix}
0 & -\frac{1}{2} & 0
\vspace{1mm}\\
-1 & -\frac{N_c}{4} & -\frac{N_c^2-4}{4N_c}
\vspace{1mm}\\
0 & -\frac{N_c}{4} & -\frac{N_c}{4}
\end{pmatrix}
, \nonumber
\\
\bm{T}_2 \cdot \bm{T}_3 = \bm{T}_1 \cdot \bm{T}_4 &=
\begin{pmatrix}
0 & \frac{1}{2} & 0
\vspace{1mm}\\
1 & -\frac{N_c}{4} & \frac{N_c^2-4}{4N_c}
\vspace{1mm}\\
0 & \frac{N_c}{4} & -\frac{N_c}{4}
\end{pmatrix}
.
\end{align}
Plugging the above into the factorization formula, we are able to reproduce the results of \cite{Czakon:2007ej, Czakon:2007wk} from the purely massless amplitudes of \cite{Anastasiou:2000kg, Anastasiou:2001sv}, including all the $n_h$-related terms. This provides a strong check on our factorization formula and the expressions of the $\zcal$-factors and the soft function.

\section{Summary and outlook}

In summary, in this work we have proposed a generic factorization formula for scattering amplitudes up to two loops in the high-energy boosted limit. Our formula completes that of \cite{Mitov:2006xs} by incorporating the contributions from massive loops (the $n_h$-contributions). The formula involves a soft function as an operator in color space, as well as modified $\zcal$-factors with $n_h$-dependence. Using the method of regions with analytic regulators for the rapidity divergences, we derive these new ingredients through explicit calculations. We verify our results using various form factors. In particular, we have computed two form factors that were not available in the literature: the massive loop contributions to the quark-quark vector form factor and to the gluon-gluon scalar form factor. Furthermore, the scattering amplitudes for $t\bar{t}$ production provide a strong validation of our results, with non-trivial color structures.

Our factorization formula can be used to obtain approximate expressions for complicated two-loop massive amplitudes from the corresponding much simpler massless ones. While such an approximation is only valid in the high-energy limit, it can be combined with low-energy approximations (e.g., in the threshold limit or the soft limit) to get reasonable results in the whole phase space. This may find applications in cutting-edge problems such as the NNLO QCD corrections to $t\bar{t}Z$, $t\bar{t}H$ $t\bar{t}j$ and $t\bar{t}t\bar{t}$ production processes. Combined with suitable RGEs, our formula can also be used to resum the large mass logarithms to all orders in the coupling constant, either in the high-energy boosted limit alone, or in the double boosted/threshold limit considered in \cite{Ferroglia:2012ku, Ferroglia:2013awa, Pecjak:2016nee, Czakon:2018nun}.

Finally, it will be interesting to think about the extension of the factorization formula to three loops and beyond. In particular, it is worth asking whether the three-loop soft function involves correlations among three external legs, similar to the two-loop anomalous dimension governing the IR divergences of massive amplitudes, and how to determine this kind of contributions through explicit calculations. This will be left for future investigations.

\acknowledgments
	Li-Lin Yang would like to thank Andrea Ferroglia and Ben Pecjak for earlier collaborations on related subjects.
	This work was supported in part by the National Natural Science Foundation of China under Grant No. 12375097, 11975030 and 12147103, and the Fundamental Research Funds for the Central Universities.

\appendix

\section{Renormalization constants}
\label{sec:ren_const}

In this Appendix we collect the renormalization constants relevant to our calculations. All renormalization constants are expanded in terms of the renormalized strong coupling $\alpha_s$. The renormalization constants for the strong coupling and the $Hgg$ effective coupling $\lambda$ are given by
\begin{align}
Z_{\alpha_s} &= 1 - \left( \frac{\alpha_s}{4\pi} \right) \frac{\beta_0}{\epsilon} + \left( \frac{\alpha_s}{4\pi} \right)^2 \left( \frac{\beta_0^2}{\epsilon^2} - \frac{\beta_1}{2\epsilon} \right) , \nonumber
\\
Z_{\lambda } &= 1 - \left(\frac{\alpha _s}{4 \pi }\right)\frac{\beta _0}{\epsilon }+\left(\frac{\alpha _s}{4 \pi }\right)^2\left(\frac{\beta _0^2}{\epsilon ^2}-\frac{\beta _1}{\epsilon }\right) ,
\end{align}
where the $\beta$-function coefficients are
\begin{align}
\beta_0 &= \frac{11}{3} C_A + \frac{4}{3} T_F n_f \,, \nonumber
\\
\beta_1 &= \frac{34}{3} C_A^2 - \frac{20}{3}C_A T_F n_f - 4 C_F T_F n_f \,.
\end{align}
The renormalization constant for the light-quark field is
\begin{align}
Z_q &= 1 + \left( \frac{\alpha_s}{4\pi} \right)^2 C_F T_F \sum_h \left( \frac{\mu^2}{m_h^2} \right)^{2\epsilon} \left( \frac{1}{\epsilon} - \frac{5}{6} \right) . \nonumber
\end{align}
The two-loop renormalization constant for the massive quark field with mass $m_Q$ arising from a massive quark loop with mass $m_h$ is given by
\begin{align}
Z_{Q,n_h}^{(2)} &= C_F T_F \Bigg\{ \left[\frac{1}{\epsilon}\left( 1 + 4\ln\frac{\mu^2}{m_Q^2} \right) + 6\ln^2\frac{\mu^2}{m_Q^2} + \frac{22}{3}\ln\frac{\mu^2}{m_Q^2} + \frac{947}{18} - 30\zeta_2 \right]  \nonumber \\
& + \sum_{h\neq Q} \Bigg[\frac{1}{\epsilon}\left( 1 + 4\ln\frac{\mu^2}{m_Q^2} -8H(0,x) \right) + 6\ln^2\frac{\mu^2}{m_Q^2} + \frac{22}{3}\ln\frac{\mu^2}{m_Q^2} + \frac{443}{18} \nonumber \\
& + \left(10 -36x -60x^3 + 24x^4\right)\zeta_2 + 28x^2 + \left(\frac{32}{3} - 16\ln\frac{\mu^2}{m_Q^2} +16x^2\right)H(0,x)  \nonumber \\
& - \left(8 + 12x + 20x^3 + 24x^4\right)H(-1,0,x) +  \left(8 - 12x - 20x^3 + 24x^4\right)H(1,0,x)  \nonumber \\
& + \left(32 +  48x^4\right)H(0,0,x) \Bigg] \Bigg\} \,,
\end{align}
where $x = m_h/m_Q$.
For the gluon field renormalization, we only give the terms with a single massive quark loop (i.e., we drop the $n_h^2$ and the $n_h n_l$ terms):
\begin{align}
	Z_g^{(1),h} &= T_F \biggl[-\frac{4}{3 \epsilon }-\frac{4 L_{m}}{3}-\epsilon  \left(\frac{2 L_{m}^2}{3}+\frac{2 \zeta _2}{3}\right)+\epsilon ^2 \left(\frac{4 \zeta _3}{9}-\frac{2 L_{m}^3}{9}-\frac{2 \zeta _2 L_m}{3}\right)\biggl] \,, \nonumber \\
	Z_g^{(2),h} &= T_F \biggl\{ C_A\biggl[\frac{35}{9 \epsilon ^2}+\frac{1}{\epsilon }\left(\frac{26 L_{m}}{9}-\frac{5}{2}\right)+\frac{4 L_{m}^2}{9}-5 L_{m}+\frac{13 \zeta _2}{9}+\frac{13}{12}\biggl]\nonumber \\
		      &\qquad+C_F\left(-\frac{2}{\epsilon }-4 L_{m}-15\right)\biggl\} \,,
\end{align}
where $L_m=\ln(\mu^2/m_h^2)$.

\bibliographystyle{JHEP}
\bibliography{references_inspire.bib}

\end{document}